\documentclass[12pt]{article}
\usepackage{putex}
\usepackage{graphicx}

\begin{document}

\title{The gauge-string duality and heavy ion collisions}

\authors{Steven S. Gubser}

\institution{PU}{Joseph Henry Laboratories, Princeton University, Princeton, NJ 08544}

\abstract{I review at a non-technical level the use of the gauge-string duality to study aspects of heavy ion collisions, with special emphasis on the trailing string calculation of heavy quark energy loss.  I include some brief speculations on how variants of the trailing string construction could provide a toy model of black hole formation and evaporation.  This essay is an invited contribution to ``Forty Years of String Theory'' and is aimed at philosophers and historians of science as well as physicists.}

\preprint{PUPT-2367}

\maketitle
\tableofcontents

\section{In cautious praise of string theory}

Practitioners of string theory aiming to present an overview of their subject, or to convey a flavor of it to non-specialists, have come up with some memorable expressions.\footnote{The remarks quoted in the first two paragraphs come from various sources.  ``Theory of everything'' came to early prominence in \cite{Ellis86}.  String theory's development is characterized as backward in \cite{Kaku:2005nh}.  The phrase ``string miracles'' occurs in \cite{Dine88}.  Witten's remark appeared in print in \cite{Lemonick04}.  Townsend's quote hasn't appeared in print as far as I know, and it is more properly understood as the closest paraphrase that my memory can provide.

The expression ``Not even wrong'' is apparently due to Pauli, but its application to string theory is most prominent in \cite{Woit:2006js}.  ``The Trouble with Physics'' is the title of \cite{Smolin:2006pe}.  Wilczek's quote appeared in \cite{Wilczek09}.}  String theory is (a candidate for) the ``theory of everything;'' the theory which grew backwards; a theory of ``string miracles;'' and, in Edward Witten's words, ``a bit of 21st century physics that somehow dropped into the 20th century.''  Paul Townsend remarked about ten years ago, ``Someday we will discover the ultimate theory underlying all of physics, and we will call it string theory.''

In painful contrast, critics describe string theory as ``not even wrong.''  The rise of string theory, in Lee Smolin's words, is the ``Trouble with Physics.''  Less dark, but still clearly critical, is Frank Wilczek's remark, ``String theory is sort of the extreme of non-empirical physics.''

Here is how I would trace the logic of the debate which I reduced to sound bites in the previous paragraphs.  String theorists characterize their theory as exceptional in some way.  The main claim, usually, is that string theory is uniquely positioned to solve the biggest problems in fundamental physics.  Critics of the theory seldom convince me that they are making a serious bid to refute this claim.  Instead, they attack string theory for failing to meet the ordinary standards, norms, and principles of theoretical physics.  For example, string theory is characterized by its critics as untested and excessively mathematical.  As biting as these criticisms may seem, they are (with a caveat I will come to next) attacks on a straw man.

Here is the caveat---and the main concession that I willingly make to the critics.  Their straw man is based on common understanding of what claims of exceptionalism imply.  Indeed, such claims are generally understood as implying that ordinary standards, norms, and principles need not be applied.  Of course, ordinary standards {\sl should} be applied to string theory, no matter how amazing its theoretical reach might become.  In particular, we should be worried by the degree to which string theory remains untested by experiment.  We should entertain some skepticism about the utility of formal constructions that have no foreseeable application to phenomenology.  And we should feel at least a modicum of concern about how far ahead of mathematical rigor the technical developments of string theory have run.  In my experience, these concerns are widely felt by the string theory community, even though they are seldom stated up front.  Perhaps it is precisely our reticence to articulate these concerns that gives critics an opening for their straw man argument.  In short, the aim for proponents of string theory should be to give the theory its due praise without hinting at excuses for its known weaknesses.

The main truth that I would draw out of the impressive statements I started with is that string theory sits at the center of a locus of ideas that spans a large fraction of modern theoretical physics.  String theory exerts a unique pull on the imagination of high-energy theoretical physicists---a pull now being felt by theorists of other stripes, including nuclear and condensed matter physicists.

In the remainder of this essay, I plan to discuss the particular corner of string theory that I am most interested in, namely the gauge-string duality.  More than most subfields of string theory, it possesses the two qualities I described in the previous paragraph: on one hand, it draws in the imagination; on the other, it reaches out to many other fields of theoretical physics.  In an attempt to convey these qualities, I once wrote, ``I think about how black holes in a fifth dimension relate to collisions of heavy ions, and people take me seriously.''  Readers of the remainder of this essay should be in a position to judge whether I made a meaningful claim to success while skirting the dangers of exceptionalism.

\section{The gauge-string duality}

The essential ideas of the gauge-string duality were articulated in the late 1990's by Juan Maldacena \cite{Maldacena:1997re}; by Igor Klebanov, Alexander Polyakov, and myself \cite{Gubser:1998bc}; and by Edward Witten \cite{Witten:1998qj}.  The grand idea is that quantum field theories that do not include gravity are often equivalent to string theory or M-theory (or perhaps to some other, as yet undiscovered, theory of quantum gravity) in a higher dimensional spacetime.  While I do not aim to trace the history of the gauge-string duality fully, some comparisons to the earlier ideas are in order.  By the middle 1990's, string dualities abounded, so the notion of sharp equivalences between superficially very different theories was well established.  For example, the duality between type IIA string theory and M-theory \cite{Townsend:1995kk,Witten:1995ex} illustrated the possibility of a fairly comprehensive mapping (purportedly extensible to a complete mapping) between a theory in ten dimensions and a theory in eleven dimensions.  However, this mapping is an extension of Kaluza-Klein theory, and so rather different from the gauge-string duality.

A more direct progenitor of the gauge-string duality was the extensive study of black hole microstates from D-brane constructions, initiated by Strominger and Vafa \cite{Strominger:1996sh}.  The big successes of this program hinged upon finding $1+1$-dimensional conformal field theories (CFTs) which could account for the Bekenstein-Hawking entropy of black holes as well as explaining the interaction of black holes with low-energy probes \cite{Das:1996wn}.  It was clear that the CFTs effectively stood in lieu of the black hole horizon, at least for the purposes of specific calculations.  This replacement of a horizon by a field theory provided an interesting link to ideas of holography that had been proposed earlier by 't Hooft and Susskind \cite{tHooft:1993gx,Susskind:1994vu}.  The holographic paradigm hinges on the idea that in a theory of quantum gravity, the microstates do not occupy the bulk of spacetime, but instead, in some appropriate sense, reside on boundaries.  Black holes provide a canonical example: In the spirit of John Wheeler's ``It from bit,'' the entropy of a horizon is considered to arise from degrees of freedom that reside on the horizon.  The membrane paradigm \cite{Thorne:1986iy} provides a calculational framework based on essentially this point of view.

In the gauge-string duality, the holographic principle finds a precise and powerful expression, thanks to the conformal properties of the spacetimes in which the gravitational theory resides.  The essential property of these spacetimes is that they have a natural boundary.  This boundary is not the horizon.  It is a boundary on which null geodesics begin and/or end.  The field theory to which the gravitational theory is dual can be defined on that boundary.  At first the presence of the boundary seems like a complication, because to specify the forward time evolution of the gravitational theory, even classically, one has to have precise information about the boundary conditions, not just on an initial time-slice, but also on the boundary where null geodesics begin and end.  But it turns out that this is just what one needs in order to have a complete mapping between the gravitational theory and the dual field theory.  The boundary conditions one needs on the gravitational side are in perfect correspondence to all the ways one can deform the field theory.  What you do with the field theory tells the gravitational theory what boundary conditions it must satisfy.  In the classical limit, the gravitational theory finds the optimal configuration satisfying specified boundary conditions, and the action of that optimal configuration corresponds to the generator of connected Green's functions of the field theory.  If one moves beyond the classical limit on the gravity side, a more generally valid phrasing is that the partition functions of the gravity theory and the field theory coincide when the former is subjected to boundary conditions corresponding to the deformations of the latter.  On one hand, this prescription is deeply in accord with the holographic principle, since all the degrees of freedom of the bulk theory are encoded on the boundary.  On the other hand, it privileges the boundary of spacetime over more {\it ad hoc} boundaries, leading to a more precise and complete statement.

The equivalence of gravity in the bulk of spacetime with field theory on the boundary subsumes the statements of semi-classical black hole thermodynamics which have been a touchstone for theoretical work in gravity for decades: When there is a black hole in the bulk, its Hawking temperature is the temperature of the field theory; its Bekenstein-Hawking entropy is the entropy of the field theory; and its action is (up to regularization and a factor of the temperature) the free energy of the field theory.

The reason to expect an equivalence along the lines of the gauge-string duality is that D-branes (or, more precisely, large stacks of D-branes) admit alternative descriptions in terms of open and closed strings.  The low-energy dynamics of open strings with their ends attached to D-branes is a non-gravitational field theory.  The reason this is so, roughly speaking, is that the open string spectrum includes massless degrees of freedom up to spin $1$, but not a graviton.  On the other hand, at least for some special cases (notably the D3-brane and the D1-D5 bound state), the geometry of many coincident branes is smooth and free of singularities: in a sense, the branes ``disappear'' down a warped throat, much in the way that electric charge disappears down the throat of an extremal Reissner-Nordstrom black hole.  This geometrical description of many coincident branes is a closed string description, because it is the closed string sector which includes the graviton, as well as the generalized gauge fields which carry the charges of the branes.

The gauge-string duality can be understood, at least in part, as the necessary equivalence between two ways of describing the same physics---namely, the low-energy dynamics of a large number of coincident branes.  More broadly, much of the progress in string theory since 1995 can be characterized as advances in understanding localized defects in spacetimes: singularities and their resolutions, branes and intersections of branes.  Part of the appeal of this line of work is that the dynamics of localized defects in string theory is extraordinarily rich and interesting, encompassing non-abelian gauge theories, non-perturbative dynamics, and renormalization group flows---all subjects of great interest independent of string theory.

Equivalence between gravitational theories in spacetimes with boundaries and non-gravi\-tational theories that can be defined on those boundaries is quite a general idea, and part of its charm is that it does not appear to depend particularly on supersymmetry or other special features of string theory.  However, symmetries---especially relativistic conformal symmetry---play a crucial role in the best understood examples.  The relativistic conformal group is an extension of the Lorentz group.  In addition to boosts and rotations, it includes translations, dilations, and so-called special conformal transformations which preserve the causal structure of spacetime but rescale spacetime volumes in a more complicated way than dilations.  In four dimensions, where the Lorentz group is $SO(3,1)$, the conformal group is $SO(4,2)$.  Relativistic quantum field theories with vanishing beta function are (up to possible anomalies) invariant under conformal transformations, and are termed conformal field theories.  The gravity duals of CFTs are gravitational theories in five-dimensional anti-de Sitter space, which is denoted $AdS_5$.  This curved spacetime can be realized as an $SO(4,2)$-symmetric hyperboloid embedded in flat ${\bf R}^{4,2}$.  It inherits the $SO(4,2)$ symmetry of ${\bf R}^{4,2}$ in the same way that the sphere inherits the $SO(3)$ rotational symmetry of three-dimensional flat space ${\bf R}^3$.  Relativistic conformal symmetry has played such an  important role in the development of the gauge-string duality that the duality itself is commonly referred to as AdS/CFT, short for the anti-de Sitter / Conformal Field Theory correspondence.

In attempts to apply the gauge-string duality to other areas of physics (notably nuclear and condensed matter physics) an assumption is often made early on that conformal invariance is a reasonably good symmetry of the physical system under consideration.  This is clearly a tricky assumption for nuclear physics, where the phenomenon of confinement picks out a particular energy scale.  If anything, it is more fraught with danger in a condensed matter setting, where the emergence of relativistic conformal invariance in strongly correlated many-body systems is rare.  In the next two sections, I will discuss applications to nuclear physics at greater length and explain how, in particular examples, interesting comparisons can nevertheless be made between the physics of the quark-gluon plasma and a theory with relativistic conformal symmetry.

\section{String theory and heavy ion collisions}

In 2003, a simple and striking calculation was published by Policastro, Son, and Starinets \cite{Policastro:2001yc}: Using the gauge-string duality, they showed that in the strong coupling limit, and at finite temperature, the shear viscosity of ${\cal N}=4$ super-Yang-Mills theory was ${1 \over 4\pi}$ times the entropy density.  This calculation was the beginning of a serious effort to relate string theory constructions to heavy ion physics.\footnote{It is striking that early explorations \cite{Damour79} of what became the membrane paradigm of black hole horizons revealed essentially the same result for the shear viscosity as found in \cite{Policastro:2001yc}.  But the interpretation of that early work is obscured by the existence of a negative (and so apparently unphysical) bulk viscosity.  Also, the horizons considered in \cite{Damour79} are of finite extent, which obstructs the hydrodynamic limit.  In any case, the context of the gauge-string duality, together with the motivation from heavy-ion physics to be explained below, make clear the proper physical interpretation of the result, as well as its importance.}  To explain why the calculation was important and why the subsequent effort gained such traction, both among string theorists and nuclear theorists, I should first explain how ${\cal N}=4$ super-Yang-Mills comes up in the gauge-string duality and also summarize the outlines of what heavy ion collisions are.

I have already noted that a key underpinning of the gauge-string duality is the study of multiple coincident D-branes in string theory.  The simplest such construction is $N$ D3-branes in otherwise flat ten-dimensional spacetime, as a configuration in type~IIB superstring theory.  The low-energy dynamics of the open strings ending on the D3-branes is the aforementioned ${\cal N}=4$ super-Yang-Mills theory.  This is a particularly simple non-gravitational quantum field theory whose beta function vanishes for symmetry reasons.  The symmetry in question is ${\cal N}=4$ supersymmetry, which is the most supersymmetry a non-gravitational field theory can have in four dimensions.  The key ingredient in the dynamics of all Yang-Mills theories is the spin $1$ gluon field, which in field theory terms owes its existence to a gauge symmetry, which is $SU(N)$ for $N$ D3-branes.  Felicitously, gluons associated with $SU(3)$ gauge symmetry underly quantum chromodynamics (hereafter QCD), which is the theory that describes most of what goes on inside the atomic nucleus, and also most of what happens in a heavy ion collision.  There are profound differences between ${\cal N}=4$ super-Yang-Mills theory (hereafter SYM or ${\cal N}=4$ SYM) and QCD, the most striking being the lack of any supersymmetry in QCD and the existence of a negative beta-function which leads to confinement.  In contrast, SYM possesses full relativistic conformal invariance, which means that no scale is privileged, and in particular confinement is impossible.  Conformal invariance is a {\it sine qua non} for a field theory to have an $AdS_5$ dual, and ${\cal N}=4$ SYM was the original example of the duality.  Its dual is more properly $AdS_5 \times S^5$, which is the near-horizon geometry of a large number of D3-branes.  This geometry is weakly curved (meaning that methods based on elaborations of Einstein gravity can be used, instead of calling on the full type~IIB string theory) provided $N \gg 1$ and that the 't Hooft coupling is strong.  The 't Hooft coupling is a measure of the probability for a gluon to split into two gluons, and it generally controls the strength of all interactions in ${\cal N}=4$ SYM.  QCD also is characterized by a 't Hooft coupling, but it changes with energy scale according to the renormalization group equations, on account of the non-vanishing beta function.  In particular, it becomes large as one approaches confinement.  Strong coupling in reference to gauge theories generally refers to a large value of the 't Hooft coupling.

Let's recap.  Both ${\cal N}=4$ SYM and QCD have gluons, but their other ingredients and properties differ.  The gauge-string duality provides a tool for understanding the strong coupling limit of SYM.  Qualitative properties of SYM at strong coupling---like the small value of the shear viscosity compared to the entropy density---might be expected to carry over to QCD at energy scales not too far above the confinement scale.  Such expectations clearly hinge on a degree of optimism that can be called into question.  Perhaps it is fair to say that the burden of proof on the question whether a quantitative relation between SYM and QCD exists falls more on the proponents than the skeptics.

Now, let's pass on to high-energy collisions.  Particle physicists have long been eager to learn everything possible from high-energy collisions, but the favorites have been collisions of electrons and positrons, electrons and protons, and protons and anti-protons.\footnote{Collisions of protons on protons are also common, and the LHC focuses on this type of collision.}  It is not hard to see why from a theoretical point of view.  The simpler the objects one collides, the more control one has in understanding the initial state.  Also, the interactions between electrons and positrons are electroweak, and therefore the scattering amplitudes are small, making a perturbative approach feasible.  Collisions of protons and anti-protons are already complicated compared to electrons and positrons: so much so that Leon Lederman joked about how it was like colliding pails of garbage and waiting to see a pearl drop out \cite{Lederman97}.  Nevertheless, perturbative techniques, supplemented by renormalization group methods, are quite successful in describing these collisions, particularly the subset where a large momentum is transferred during the scattering event.

In heavy ion collisions, one strips all the electrons off of a heavy isotope (gold and lead are the typical favorites) and collides the resulting ions at relative velocities comparable to those reached in proton-anti-proton collisions.  The most authoritative experimental reviews to date are \cite{Adcox:2004mh,Arsene:2004fa,Back:2004je,Adams:2005dq}, and a wide cross-section of theoretical perspectives is presented in \cite{Abreu:2007kv}.  Describing heavy ion collisions in perturbative quantum field theory seems like a hopeless task because there are too many interactions.\footnote{Nevertheless there is a robust effort to describe many facets of the initial state and early stages of the collision using a combination of perturbative and semi-classical methods.  In addition to \cite{Abreu:2007kv}, see for example the reviews \cite{McLerran:2008uj,Gelis:2010nm}.}  One can hope that pushing far away from the perturbative regime will eventually reveal some new regime of simplicity.  That in fact happens, and the new regime of simplicity is fluid dynamics.  Fluid dynamics relies on frequent interactions among constituent particles: so frequent that it is essentially impossible to keep track of them individually, and meaningful instead to describe their collective behavior in terms of local density and velocity.  There is a vigorous and remarkably successful enterprise to describe bulk features of heavy-ion collisions in terms of relativistic fluid dynamics: for a pedagogical review, see for example \cite{Heinz:2004qz}.  It is worth noting that fluid dynamical descriptions of the atomic nucleus have an old and honorable tradition, dating back to the liquid drop model of the nucleus used by Bohr and Wheeler to describe nuclear fission \cite{Bohr:1939ej}.  In that historic work, the fluid elements were understood to be individual nucleons (protons and neutrons).  In heavy-ion collisions, the nucleons are utterly destroyed, and the fluid that emerges is composed of strongly interacting quarks and gluons.  It is called the quark-gluon plasma, and it is by far the hottest substance created in the laboratory: Recent measurements suggest a temperature as high as $4$ trillion degrees Kelvin \cite{Adare:2009qk}.

The whole enterprise of comparing string theory calculations with heavy-ion collisions hinges on the description of the quark-gluon plasma---or a thermal state with comparable properties---in terms of a dual black hole.  A favorite starting point is a black hole solution known as $AdS_5$-Schwarzschild, which captures the strong coupling dynamics of finite-temperature ${\cal N}=4$ SYM theory \cite{Gubser:1996de}.  In the remainder of this section I will discuss two particularly successful calculations in this framework.  The first is the shear viscosity computation mentioned at the beginning of this section.  The second is a computation of the rate of energy loss of energetic particles traversing the thermal medium.

Experimental observations of the anisotropic expansion of nuclear matter in off-center heavy-ion collisions prompted hydrodynamic analyses which show consistency with the data provided the shear viscosity is much smaller than the entropy density: see for example \cite{Luzum:2008cw}.  A rough understanding of the phenomenon of anisotropic expansion can be achieved as follows.  Consider an initial lump of interacting quarks and gluons formed in the collision.  Usually this lump will not be symmetrical.  If the quarks and gluons interact very little, then most of what happens during the expansion is that these particles free-stream away from one another until they go through the hadronization process that turns them into observable particles.  This free-streaming has a strong tendency to erase initial anisotropies.  On the other hand, if the quarks and gluons interact very strongly, the initial lump will wobble and shake as it expands, and the imprint of its initial anisotropies will remain in its collective outward flow.  This imprint is more visible when shear viscosity is small, because viscosity generally has a damping property that diffusively erases irregularities.  Shear viscosity is hard to compute reliably in the framework of perturbative quantum field theory, and existing estimates \cite{Arnold:2000dr} tended to put the shear viscosity at unacceptably high values.  So the calculation of \cite{Policastro:2001yc} was phenomenologically attractive.  Moreover, it was subsequently shown in \cite{Kovtun:2004de}, among other works, to be quite robust in the sense that alterations of the gravitational theory changed it either by small corrections, or in many cases not at all.

Another key aspect of heavy-ion phenomenology is the suppression of high-energy particles in the final state, relative to collisions of protons.  The common understanding of this suppression hinges on a thought experiment.  Instead of colliding two gold nuclei, imagine separating the constituent nucleons of each nucleus just a little in the direction of their motion.  Then, instead of one big collision, there is a series of pairwise nucleon collisions.  There are accepted ways of estimating how many such pairwise collisions occur, and individual proton-proton collisions at the relevant energies are experimentally well studied.  So there is a definite prediction, starting from data on proton-proton collisions, of the spectrum of particles that emerge in a heavy ion collision, based on replacing the heavy ion collision by an appropriate number of pairwise nucleon collisions.  This whole approach is recognized as providing only baseline expectations whose success or failure help teach us about what really happens in heavy ion collisions.  Suppression of high-energy particles is a case in point.  A characteristic feature of QCD is that some fraction of the final state particles in proton-proton collisions are quite energetic.  If a heavy ion collision were really the same as a series of nucleon-nucleon collisions, then the fraction of final state particles above a specified energy would be the same as in proton-proton collisions.  Instead, considerably fewer particles are highly energetic.  According to common understanding, the initial stages of a heavy ion collision are essentially identical to the series of pairwise nucleon collisions discussed above; but as particles try to escape toward the detectors, the detritus of other pairwise nucleon collisions gets in their way and saps their energy.  The detritus is understood to be the quark-gluon plasma, whose fluid properties I discussed above.  Thus the suppression of high-energy particles in the final state of heavy ion collisions is understood as an opportunity to probe the properties of the quark-gluon plasma.

For comparison with experimental data on suppression of high-energy particles (well reviewed in \cite{Adcox:2004mh,Arsene:2004fa,Back:2004je,Adams:2005dq}), a key quantity is the typical length that an energetic particle can travel through the quark-gluon plasma before it loses a large fraction of its original energy.  Experimental evidence suggests that this length is only a few times the proton radius.  (The proton radius is a little less than $1\,{\rm fm}$.)  So again we are confronted with a situation which is far from the free-streaming picture where perturbative quantum field theory is best justified.\footnote{I would be remiss not to note that there are QCD calculations, both in perturbative and partially perturbative frameworks, which have made impressive headway on the problem of describing rapid energy loss.  For an introductory review of these methods, see for example \cite{CasalderreySolana:2007pr}.}

In 2006, calculations appeared in string theory which in various ways probe the rate of energy loss by energetic particles in a strongly coupled thermal medium.  These calculations were performed by Rajagopal, Liu, and Wiedemann \cite{Liu:2006ug}; by Herzog et.\ al.\ \cite{Herzog:2006gh}; by Casalderrey-Solana and Teaney \cite{Casalderrey-Solana:2006rq}; and by the author \cite{Gubser:2006bz}.  While there are some tensions among these calculations, they generally point toward rapid energy loss.  Two obstacles to leveraging these string theory calculations into quantitative predictions for QCD are 1) ${\cal N}=4$ SYM has about three times as many degrees of freedom as QCD; and 2) there is some uncertainty on how best to choose the gauge coupling in ${\cal N}=4$ SYM.  I attempted a resolution of these two points through a comparison of static properties of quark-anti-quark pairs between lattice QCD and ${\cal N}=4$ SYM, with the result that charm quarks propagate about $2\,{\rm fm}$ before losing a large fraction of their energy.  A phenomenological study \cite{Akamatsu:2008ge} indicates that this estimate is consistent with data.

Low shear viscosity and rapid energy loss are two of the high water marks of the effort to relate string theory to heavy ion collisions.  It is perhaps appropriate to close this section with some reflections on their relative merits.  The shear viscosity calculation has special appeal because of its universality and simplicity.  Macroscopic liquids have shear viscosity and entropy density, and much has been made of the observation that their ratio (appropriately defined to avoid difficulties with superfluids) is always greater than the $1/4\pi$ that comes up in the study of black hole horizons.  A significant effort has emerged aimed at tracking more precisely how horizons relate to fluid dynamics.  On the other hand, the heavy ion data constrains the shear viscosity mostly in the direction of an upper bound.  It is possible, at least in terms of the data, that the true shear viscosity of the quark-gluon plasma is significantly less than $1/4\pi$ times the entropy density.  Much depends on the initial conditions for hydrodynamic evolution, in particular the thermalization time.

The energy loss calculations are less universal and harder to relate to macroscopic systems (see however \cite{Gubser:2009qf}).  Moreover, the key results always involve the gauge coupling, and there is no punchline quite as simple as the $1/4\pi$ result in the shear viscosity story.  And yet the experimental situation is, if anything, better, in that both upper and lower bounds on the rate of energy loss are supported by the data, and the results from string theory, at least as applied in \cite{Akamatsu:2008ge}, lie between the two.  Altogether, the shear viscosity and drag force calculations in the gauge-string duality are best viewed as two sides of the same coin: they both emphasize a wholly non-perturbative description of the quark-gluon plasma based on a black hole horizon, and they both point toward dynamics that is far from the free-streaming limit.

In the next section I will explain more fully (though still qualitatively) how the string theoretic calculation of energy loss is done.  While the chance for a comparison with data motivated this calculation, it has some interesting and surprising theoretical features related to a causal horizon on the string worldsheet.

\section{The trailing string}

In the early literature of the gauge-string duality, the attractive force between static quarks and anti-quarks was examined \cite{Rey:1998ik,Maldacena:1998im}.  In a conformal field theory at zero temperature, that force depends inversely on the square of the separation between the quark and anti-quark.  It arises on the gravity side because a fundamental string comes out of the quark, goes down into the boundary, and then goes across and up again to the anti-quark.  A purely classical calculation on the string theory side, based on the area of the string worldsheet, leads to a definite form for the attractive force.  This calculation is easily extended to the case of finite temperature, where the attractive force weakens compared to the zero-temperature result due to the influence of the black hole.

The force between a static quark and anti-quark is one of the standard quantities studied in lattice QCD: for a concise introduction, see for example \cite{Karsch:2006sf}.  It is a very revealing quantity, as it encodes both the departures from conformality due to renormalization group flow and finite temperature corrections due to Debye screening.  I proposed in \cite{Gubser:2006qh} that when comparing calculations in QCD to calculations in ${\cal N}=4$ super-Yang-Mills theory---the latter performed using the gauge-string duality---a sensible approach is to consider thermal states with the same energy density in both theories, and to set the Yang-Mills coupling in such a way that the force between a static quark and anti-quark matches at a radius significantly less than the length scale of confinement.  This approach is to an extent {\it ad hoc}, in that one might reasonably prefer comparisons at fixed temperature, and/or comparisons where the Yang-Mills coupling in ${\cal N}=4$ super-Yang-Mills theory is set equal to the gauge coupling in QCD.  However, using fixed energy density is a simple way to compensate for the fact that ${\cal N}=4$ super-Yang-Mills theory has about three times as many degrees of freedom as QCD; and using the force between quarks and anti-quarks to normalize the coupling seems particularly well motivated when considering the drag force on a single quark, as I will discuss next.

Finite-temperature lattice QCD is well adapted to quantities which can be measured in static systems, like the equation of state and the attraction between static quarks and anti-quarks.  It is much less well suited to quantities where evolution in real time is an essential feature.  This is because the lattice is constructed using a Wick-rotated spacetime, with real time replaced by periodic Euclidean time, where the period is the reciprocal of the temperature.  In the gauge-string duality, the Wick rotation between Minkowskian geometries and Euclidean geometries is, at least in many interesting circumstances, almost trivial.  So properties like the drag force on a moving quark are fairly straightforward to calculate.\footnote{Viscosity also relates to a time-dependent situation, where a fluid has some non-trivial motion (shearing flow, for example) which damps out over time at a rate that the viscosity controls.  For viscosity (as well as other transport properties) a standard approach is to extract an appropriate transport coefficient from a two-point correlator, which represents a perturbation of a uniform, static system.  In principle, lattice QCD practitioners can get at the appropriate two-point functions, and the effort to do so for viscosity has had some notable success \cite{Meyer:2007ic}.  So the point is not that lattice QCD cannot possibly get at the quantities that string theorists compute; rather, it's that string theorists have, at least temporarily, the advantage of a simpler and more nearly analytically tractable approach.}

The way the drag force calculation goes can be summarized in four main points:
 \begin{itemize}
  \item A single heavy quark at rest in a thermal bath can be represented on the gravity side as a string hanging straight down from the boundary into the horizon of $AdS_5$-Schwarzschild.  This string cannot break because the string theory we're discussing is type~IIB, where string breaking is explicitly forbidden except where there are D-branes.  The string's upper end must remain where the quark sits, and the rest of the string also remains stationary, at least in the classical limit where we ignore quantum fluctuations of the position of the string.
  \item The endpoint of the string, which is attached to the heavy quark, can be required to move along any timelike trajectory on the boundary.  The speed limit for its motion is the speed of light in the boundary theory.  The speed of light is identical in the bulk theory, but because the bulk spacetime is curved, the speed of light {\sl appears} to a boundary observer to get slower as one goes down into the bulk geometry.  This is a key point, so let's be quantitative.  If $t$ is time of the boundary theory, $x$ is one of the spatial directions, and $r$ is the fifth dimension of $AdS_5$, then a light ray moving parallel to the boundary at a ``depth'' $r$ moves at an apparent speed $v_{\rm max}(r) = \sqrt{|g_{tt}(r)/g_{xx}(r)|}$.  As one approaches the black hole horizon, $v_{\rm max}(r) \to 0$.  This is an example of gravitational redshift: everything that happens near a black hole horizon seems to distant observers to occur very slowly.
  \item Suppose the heavy quark moves at a constant velocity with magnitude $v_0 \in (0,1)$, where for convenience I have now set the speed of light to unity.  Sufficiently close to the horizon, we will find $v_{\rm max}(r) < v_0$.  This means that part of the string (the part close to the horizon) cannot ``keep up'' with the heavy quark.  So the string trails out behind the heavy quark.  Its precise shape can be determined by minimizing the Nambu-Goto action \cite{Herzog:2006gh,Gubser:2006bz}.  The trailing string exerts a backward force on the heavy quark.  This is the drag force.  A closely related calculation of stochastic forces on the string appeared independently \cite{Casalderrey-Solana:2006rq}.  
  \item The final result for the drag force indicates that it is proportional to the tension of the fundamental string.  This string tension appears in the same way in the force between static quarks.  So what string theory offers, in practical terms, is not so much an absolute calculation of the drag force on a single quark, but a ratio of the drag force to the force between a static quark and anti-quark.  Because the latter is well studied in lattice QCD, one can put the string theory calculation together with lattice results to get a real prediction.  This, in a nutshell, is how I arrived \cite{Gubser:2006qh} at the estimate that charm quarks propagate about $2\,{\rm fm}$ before losing a large fraction of their energy.
 \end{itemize}

The trailing string has an interesting causal structure.  Recall that for any velocity $v_0 \in (0,1)$ of its endpoint, there is a part of the trailing string near the black hole horizon which cannot keep up with the endpoint.  The endpoint of the string is causally inaccessible to this part of the string, at least if one is limited to signals which propagate along the string worldsheet.  There is a horizon on the string which demarcates the region which has no causal access to the endpoint.  This horizon emits fluctuations which propagate up to the endpoint of the string.  These fluctuations are precisely analogous to the Hawking radiation from a black hole.  In the description of the dynamics of heavy quarks, they play a key role: they provide the stochastic forces which are generally expected to arise in conjunction with the dissipative drag force \cite{Casalderrey-Solana:2006rq,Gubser:2006nz,Casalderrey-Solana:2007qw}.

In the original example of the trailing string, where the background geometry is $AdS_5$-Schwarzschild, there is a horizon on the string worldsheet even in the limit $v_0 \to 0$: It coincides with the horizon of the background geometry in this limit.  However, in \cite{Gubser:2009qf}, it was observed that trailing string configurations exist even in background geometries without a horizon.  The example studied explicitly was a solution of type~IIB string theory found in \cite{Gubser:2009qm}, where $\sqrt{|g_{tt}(r)/g_{xx}(r)|}$ varies over a range $(v_*,1)$ in the bulk geometry.  The quantity $v_*$ can be understood in field theory terms as the reciprocal of the index of refraction of the zero-temperature state of matter dual to the geometry found in \cite{Gubser:2009qm}.

When the endpoint of the string travels at a speed $v_0$ which is less than $v_*$, the equilibrium shape of the string has no horizon.  If $v_0>v_*$, then there is a horizon.  This situation offers a simple venue in which to explore some of the mysteries of black hole physics.  Suppose we start with a stationary string, so that there is no horizon.  Next let's put the endpoint in motion at a velocity $v_0>v_*$.  In some appropriate sense, a horizon must eventually form on the string worldsheet.\footnote{The sense in which this must happen is that signals on the parts of the worldsheet where $v(r) < v_0$ have no choice but to propagate further away from the string endpoint.  This is a simple version of the notion of an apparent horizon.}  The horizon will produce Hawking radiation which results in stochastic forces on the string endpoint.  At a later time, we could choose to have the endpoint slow down or stop.  If we wait a sufficiently long time, any given part of the string (at any prespecified value of $r$) will regain causal access to the endpoint of the string.  So there is no horizon anymore.\footnote{There is a significant caveat to this statement: one could imagine a situation where for arbitrarily late times, there is always some part of the string worldsheet, however small it may be, which cannot send a signal to the endpoint of the string.  Such a situation would be like a black hole remnant.  To rule out such a possibility, one could impose the constraint that the string ends on a brane which is deep down in the bulk.}  In short, just by shaking the endpoint of the string, we can get a horizon to form, radiate, and then disappear.  At the end of the day, the entire string has causal access to the endpoint, so no information can be lost.

The narrative of the previous paragraph could, at least in principle, be supplemented by precise mathematics.  The string worldsheet can be treated through methods of $1+1$-dimensional conformal field theory.  Admittedly, the conformal field theory for the background geometry of \cite{Gubser:2009qm} is a complicated one, involving Ramond-Ramond fields.  But this is not a problem of principle.  Once boundary conditions on the endpoint of the string are specified, one can in principle track the wave-function of the string throughout its evolution.  This wave-function does not involve quantum gravity in any way: in the probe approximation, where the string doesn't back-react significantly on the background geometry, the wave-function would only describe the position of the string in a fixed background.  And yet, the wave-function would explicitly include a horizon, an analog of Hawking radiation, and even black hole evaporation.  Details are left as an exercise for the ambitious reader.

\section{Summary}

The gauge-string duality, conceived in contemplation of stacks of many coincident D-branes, has turned out to be a generally useful tool in exploring strongly coupled systems.  Its strengths include ease of calculation, geometrical intuitiveness, and direct relation to interesting strongly coupled gauge theories.  The most interesting example of a strongly coupled gauge theory that Nature has, to date, furnished us, is quantum chromodynamics; so it is not surprising that extensive efforts have been devoted to using the gauge-string duality to elucidate properties of QCD.  I have focused in this essay on properties related to finite temperature and heavy-ion collisions.  This is an easy choice for me since I have been working hard lately on this type of application of the gauge-string duality.  There are also extensive efforts to understand zero-temperature properties of QCD starting from AdS/CFT.  A more balanced review of these efforts can be found in \cite{Gubser:2009md}.

A special charm of studying finite-temperature effects in AdS/CFT is that one is constantly working with black hole horizons.  Viscosity and drag force are only two examples of the quantities one can calculate in this framework.  Another quantity of phenomenological interest is the characteristic time scale on which a black hole equilibrates, corresponding to the thermalization time of the dual field theory.  A crude estimate of the thermalization time is to ask how long it would take a graviton to propagate in pure $AdS_5$ down to the depth where a horizon eventually forms, and then back up to the boundary.  This leads to the estimate $\tau = {2 \over \pi T}$ for the thermalization time.  Interestingly enough, recent numerical simulations of classical black hole formation in $AdS_5$ \cite{Chesler:2010bi} bear out the validity of this estimate in a setting not dissimilar to heavy-ion collisions.  In perturbative treatments, the thermalization time typically is enhanced by a negative power of the coupling.  Such treatments generally have trouble explaining why the quark-gluon plasma equilibrates in a real heavy-ion collision before the medium blows itself apart.  Estimating $\tau = {2 \over \pi T}$, where $T$ is the peak equilibrated temperature, seems to be at least in the right ballpark for a phenomenologically acceptable account.

Wonderful as the gauge-string duality is, there are fairly sharp limitations on the types of field theories one can study, especially when one restricts to the supergravity approximation in the bulk.  The theories must have some sort of large $N$ approximation, and the examples which can be derived explicitly from string theory are usually either supersymmetric or closely related to supersymmetric theories.  The question naturally arises, what general lessons can we extract from these examples about strongly coupled phenomena?  Of course, one well-appreciated theme is that symmetries constrain dynamics in important ways.  I have tried to argue in recent papers \cite{Gubser:2010ze,Gubser:2010ui,Gubser:2011qv} that the approximate conformal symmetry of QCD above the confinement scale has not yet been fully exploited.  Another theme, which perhaps deserves a more cautious treatment, is that strongly coupled phenomena tend to be universal.  The universality of the $1/4\pi$ ratio of shear viscosity to entropy density over a broad range of black hole constructions is one example.  I would be eager to see some more general field theory understanding of how to make semi-quantitative statements about field theories in which no weakly coupled quasi-particle description is available.

\bibliographystyle{ssg}
\bibliography{essay}

\end{document}